\documentclass[]{aastex62}

\usepackage{comment}
\usepackage{lineno}
\usepackage{amsmath}
\begin{document}

\title{Multiscale Solar Wind Turbulence Properties inside and near Switchbacks measured by Parker Solar Probe}

\correspondingauthor{M. M. Martinovi\'c}
\email{mmartinovic@arizona.edu}

\author[0000-0002-7365-0472]{Mihailo M. Martinovi\'c}
\affiliation{Lunar and Planetary Laboratory, University of Arizona, Tucson, AZ 85721, USA.}
\affiliation{LESIA, Observatoire de Paris, Meudon, France.}

\author[0000-0001-6038-1923]{Kristopher G. Klein}
\affiliation{Lunar and Planetary Laboratory, University of Arizona, Tucson, AZ 85721, USA.}

\author[0000-0002-9954-4707]{Jia Huang}
\affiliation{Climate and Space Sciences and Engineering, University of Michigan, Ann Arbor, MI 48109, USA}

\author[0000-0003-4177-3328]{Benjamin D. G. Chandran}
\affil{Department of Physics \& Astronomy, University of New Hampshire, Durham, NH 03824, USA}
\affil{Space Science Center, University of New Hampshire, Durham, NH 03824, USA}

\author[0000-0002-7077-930X]{Justin C. Kasper}
\affiliation{Climate and Space Sciences and Engineering, University of Michigan, Ann Arbor, MI 48109, USA}
\affiliation{Smithsonian Astrophysical Observatory, Cambridge, MA 02138 USA.}

\author[0000-0003-1945-8460]{Emily Lichko}
\affiliation{Lunar and Planetary Laboratory, University of Arizona, Tucson, AZ 85721, USA.}

\author[0000-0002-4625-3332]{Trevor Bowen}
\affil{Space Sciences Laboratory, University of California, Berkeley, CA 94720-7450, USA}

\author[0000-0003-4529-3620]{Christopher H. K. Chen}
\affil{School of Physics and Astronomy, Queen Mary University of London, London E1 4NS, UK}

\author[0000-0002-6276-7771]{Lorenzo Matteini}
\affil{Department of Physics, Imperial College London, London SW7 2AZ, UK}

\author[0000-0002-7728-0085]{Michael Stevens}
\affil{Smithsonian Astrophysical Observatory, Cambridge, MA 02138 USA.}

\author[0000-0002-3520-4041]{Anthony W. Case}
\affiliation{Smithsonian Astrophysical Observatory, Cambridge, MA 02138 USA.}

\author[0000-0002-1989-3596]{Stuart D. Bale}
\affil{Space Sciences Laboratory, University of California, Berkeley, CA 94720-7450, USA}
\affil{School of Physics and Astronomy, Queen Mary University of London, London E1 4NS, UK}
\affil{Physics Department, University of California, Berkeley, CA 94720-7300, USA}
\affil{The Blackett Laboratory, Imperial College London, London, SW7 2AZ, UK}

\begin{abstract}

\emph{Parker Solar Probe (PSP)} routinely observes magnetic field deflections in the solar wind at distances less than 0.3 au from the Sun. These deflections are related to structures commonly called 'switchbacks' (SBs), whose origins and characteristic properties are currently debated. Here, we use a database of visually selected SB intervals---and regions of solar wind plasma measured just before and after each SB---to examine plasma parameters, turbulent spectra from inertial to dissipation scales, and intermittency effects in these intervals. We find that many features, such as perpendicular stochastic heating rates and turbulence spectral slopes are fairly similar inside and outside of SBs. However, important kinetic properties, such as the characteristic break scale between the inertial to dissipation ranges differ inside and outside these intervals, as does the level of intermittency, which is notably enhanced inside SBs and in their close proximity, most likely due to magnetic field and velocity shears observed at the edges. We conclude that the plasma inside and outside of a SB, in most of the observed cases, belongs to the same stream, and that the evolution of these structures is most likely regulated by kinetic processes, which dominate small scale structures at the SB edges. 
\end{abstract}


% -------------------- INTRODUCTION -------------------- 

\section{Introduction} 
\label{sec:Intro}

% phenomenology

\emph{Parker Solar Probe (PSP)} is the first mission to measure the inner heliosphere closer than 0.3 au from the Sun's surface, with perihelion distances of $\sim$ 0.16 au during its first three encounters with the Sun. One of the most compelling results from these first encounters is the ubiquity of so-called 'switchback' structures (SBs)\footnote{These structures are also called 'spikes', 'jets' or 'magnetic field reversals' in the literature.}, characterized by a rotation of the magnetic field $\mathbf{B}$ observed by the spacecraft \citep{Kasper_2019_Nature,Bale_2019_Nature}. Along with the magnetic field rotation, the magnitude of the solar wind bulk velocity $\mathbf{v_{sw}}$ increases by the order of the local Alfv\'en speed $v_A = B / \sqrt{\mu_0 n_p m_p}$ \citep{Matteini_2014_GRL}, where $\mu_0$ is magnetic permeability of vacuum, and $n_p$ and $m_p$ are proton density and mass, respectively. Results from previous missions had already motivated the community to invest significant effort into understanding the nature of SBs, though they were not observed nearly as frequently as by \emph{PSP}. Now, this interest has increased, as new data sets from \emph{PSP} provide significantly larger populations of SBs and more detailed information on the associated plasma and electromagnetic field conditions due to increased measurement precision and cadence. 

% first observations

\subsection{Previous observations of Switchbacks in the solar wind}
\label{sec:Intro_Observations}

The first observations of sudden magnetic field rotations were reported in \emph{Pioneer 6} observations \citep{Burlaga_1969_SoPh_B,Burlaga_1971_JGR}, and later in \emph{Helios} measurements \citep{Marsch_1981_JGR}. A decade later, measurements from the \emph{Ulysses} mission confirmed that the probability distributions of the magnetic field orientation are not uniform, but are rather skewed toward the polarity opposite to the dominant polarity of the solar hemisphere the spacecraft was observing \citep{Forsyth_1995_GRL}. This result was followed by a statistical analysis of Ulysses high latitude observations \citep{McComas_1998_GRL} that demonstrated the opposite polarity measurements occurred in a non-negligible ($\sim 7-8\%$) fraction of observations, but only a minor fraction (of the order of a percent) of these featured a complete inversion of the magnetic field \citep{Balogh_1999_GeoRL}. These events were classified as discontinuities---either tangential discontinuities (TD), usually separating two different plasma streams; or rotational discontinuities (RD), with a single stream following a folded magnetic field line \citep{Burlaga_1977_JGR,Neubauer_1981_sowi.conf}. Applying the standard Minimum Variance Analysis (MVA) procedure (see e.g. \cite{Sonnerup_1998_ISSIR}), to single-point measurements \citep{Burlaga_1969_SoPh,Mariani_1973_JGR,Smith_1973_JGR,Neugebauer_1984_JGR,Lepping_1986_JGR}, RDs are more frequently observed than TDs. In contrast, analysis of sets of manually selected events simultaneously observed by three different spacecraft at 1 au \citep{Tsurutani_1979_JGR,Horbury_2001_GRL}, and shortly after of four-point measurements from the \emph{Cluster II} mission \citep{Knetter_2003_AdSpR,Knetter_2004_JGRA}, found TDs to be dominant. These authors argued that results derived from MVA might be misleading due to presence of surface waves on TDs \citep{Denskat_1977_JGR,Hollweg_1982_JGR}, and therefore observations of magnetic field folds manifest in RDs are not as common as initial work suggested.

Although there is an evident ambiguity in the identification of SBs, they were considered to be isolated events \citep{Horbury_2001_GRL,Yamauchi_2004_JGRA,Horbury_2018_MNRAS} that had been studied in detail with some basic properties well established prior to the launch of \emph{PSP}. It was understood that the magnetic field deflection in SBs is followed by both the proton \citep{Neugebauer_2013_AIPC} and electron beam (also known as the \textit{strahl}) \citep{Kahler_1996_JGR}. Similar behavior was observed for relative drift between proton and $\alpha$ particle populations \citep{Yamauchi_2004_JGRA}. The fluctuations of the magnetic and velocity field at large scales, meaning the scales in the MHD domain with size comparable to the size of a structure, are linearly related, describing a dominantly Alfv\'enic turbulence \citep{Gosling_2009_ApJ}.
Moreover, it can be shown that the SB magnetic and velocity fluctuations are spherically polarized in the reference frame of zero electric field \citep{Gosling_2011_ApJ,Matteini_2014_GRL}, which can be well approximated by the $\alpha$ particle's velocity, due to their weak interaction with Alfv\'enic fluctuations of protons \citep{Matteini_2015_ApJ}. Some turbulent properties, e.g.
power spectra and cascade rates, are found to be very similar inside and outside of SBs, though discrete changes of velocity and the magnetic field are contributing to intermittency of turbulence within observed events \citep{Horbury_2018_MNRAS}.

Starting with the first \emph{PSP} solar encounter in November 2018, a plethora of new results regarding SBs were reported, as these structures were consistently observed and resolved with high resolution measurements for radial distances less than 0.25 au. These first results provided an overview of SB properties, finding some differences compared to previous reports. Most notably, \emph{PSP} did not observe SBs as isolated events, but rather 'clustered' phenomena \citep{DudokdeWit_2020_ApJS}, with measured parallel temperature changing as the spacecraft moves through the cluster \citep{Woodham_2020_arXiv}. This claim is supported by two independent arguments: 1) normalized magnetic field deflections from the Parker spiral (scaled to values between 0 and 1 depending on cosine value) exhibit a power law distribution, which is a property related to phenomena such as entangled filaments and flux tubes, rather than a Gaussian distribution, which would represent randomized events; and 2) Detrended Fluctuation Analysis \citep{Kantelhardt_2001_PhyA} shows an increase in long range correlations in time series for SBs compared to quiescent intervals. These 'clusters' are found to be separated by quiet periods with few SBs \citep{Horbury_2020_ApJS}. The origin of this discrepancy between \emph{PSP} and previous measurements is not easy to understand due to \emph{PSP}'s unique trajectories compared to previous missions, nearly co-rotating with the Sun's surface during significant fractions of its first encounters \citep{Bale_2019_Nature,Badman_2020_ApJS,Rouillard_2020_ApJS}. This specific trajectory is likely causing the observed magnetic field deflections to have shorter duration in measurements for larger radial distance, as the spacecraft could be slicing through an elongated structure, possibly aligned along the Parker spiral \citep{Laker_2020_arXiv}. Apart from these unexpected characteristics, many similarities between \emph{PSP} and previous SB observations remain. First, SBs are found to be dominantly Alfv\'enic structures, with direction of Poynting flux and turbulent fluctuations changing along with the magnetic field \citep{McManus_2020_ApJS,Bourouaine_2020_ApjL,Mozer_2020_ApJS}, maintaining the correlation between bulk velocity and magnetic field when the field is inverted \citep{Matteini_2014_GRL}, while proton temperature does not vary significantly through sharp field reversals \citep{Wooley_2020_MNRAS}. Second, \cite{Mozer_2020_ApJS} found that that the Poynting flux remains conserved before and after a SB, but significantly increases inside the structure. Third, the Partial Variance of Increments (PVI) of the magnetic field, a proxy for intermittency, is also found to be increased in clustered intervals \citep{Chhiber_2020_ApJS}, in accordance with previous results. Finally, it is worth emphasizing that a very careful case study by \cite{Krasnoselskikh_2020_ApJ} demonstrated that not all SBs appear to be Alfv\'enic in nature. Some of the events exhibit significant wave activity, especially at the borders of SBs. This is consistent with signatures of current sheets \citep{Farrell_2020_ApJS} and Whistler modes \citep{Agapitov_2020_ApJL}, detected at the edges of these structures, where the magnetic field is rotating. 

\subsection{A Mystery of Switchback Origins}
\label{sec:Intro_Origins}

The characteristics of SBs listed in Section \ref{sec:Intro_Observations} do not conclusively demonstrate how SB are formed or how they evolve throughout the inner heliosphere. Proposed mechanisms for the creation of SBs can be roughly classified in two groups - they either form on or very close to the solar surface, or are generated in the propagating solar wind - with both interpretations being represented in the literature. 

A connection between SB characteristics and surface source structures could lend credence to the theories suggest SBs form on or close to the surface of the Sun. Toward this end, many authors have investigated the connection between the plasma streams of increased bulk velocity and coronal jets \citep{McComas_1995_JGR,Neugebauer_1995_JGR,Culhane_2007_PASJ,Patsourakos_2008_ApJ,Neugebauer_2012_ApJ,Tian_2014_Science}\footnote{For an extensive review of the remote sensing observations of interest, see \cite{Raouafi_2016_SSRv}.}. Since the Alfv\'enic fluctuations in the chromosphere contain a significant amount of energy \citep{DePontieu_2007_Science}, one possible interpretation of our \emph{in situ} observations, also supported by extensive simulation work \citep{Uritsky_2017_ApJ,Wyper_2017_Nature,Roberts_2018_ApJ}, is that these fluctuations will continue to propagate even if a jet does not escape the Sun's gravity \citep{Horbury_2020_ApJS}. Another candidate mechanism is interchange reconnection, which arises from magnetic flux transport close to the surface through structures such as coronal loops \citep{vanBallegooijen_1998_ApJ,Fisk_2001_ApJ,Fisk_2005_ApJ}, interacting with open magnetic field lines to create S-shaped structure observed as a SB \citep{Fisk_2020ApJL,Zank_2020_ApJ}. Recent MHD simulations show that, if a SB is in equilibrium with the surrounding plasma, it can preserve its shape long enough to reach \emph{PSP} \citep{Landi_2005_ApJ,Tenerani_2020_ApJS}. Further on, it is expected to gradually merge with the surrounding plasma, most likely via parametric decay instability \citep{Derby_1978_ApJ}, a process potentially at work in the solar wind, as observed on \emph{Wind} mission \citep{Bowen_2018_ApJL}, and predicted by turbulence models \citep{Chandran_2018_JPlPh}, hybrid \citep{Matteini_2010_GeoRL} and MHD simulations \citep{Tenerani_2013_JGRA,Primavera_2019_ApJ}.

The complementary argument is that magnetic field folds are generated as the solar wind propagates. The main observational support for this argument is that the fraction of the solar wind with inverted magnetic field increases with radial distance \citep{Forsyth_1995_GRL,McComas_1998_GRL,Borovsky_2016_JGRA,Owens_2018_ApJ,Macneil_2020_MNRAS,Badman_2020_arXiv}. Also, simulations non-compressible three dimensional MHD turbulence \citep{Zhdankin_2012_ApJ} show that measured the distribution of magnetic field orientation at 1 au can be caused solely by turbulent fluctuations. Although these studies provide information on the statistics of the magnetic field and electron strahl orientation, more detailed research of strictly defined SBs, their duration, deflection angles and other parameters for purposes of effective comparison with \emph{PSP} results has yet to be performed. Regarding possible mechanisms for the generation of magnetic field reversals as the solar wind propagates outwards, MHD simulations show that they can develop from velocity shears \citep{Landi_2006_GeoRL,Ruffolo_2020_ApJ}. SBs are predicted to appear as a product of turbulent fluctuations evolving while maintaining constant magnetic field intensity \citep{Vasquez_1996_JGR}. The discontinuity, which forms due to the fact that a three dimensional configuration for which the field fluctuation is of the order of the field intensity is not expected to be possible \citep{Barnes_1976_JGR,Valentini_2019_ApJ}, creates the observational signature such the one observed by \emph{PSP} \citep{Squire_2020_ApJL}. It is important to note that, even if the role of the turbulent cascade turns out not to be crucial for SB generation, the effects of turbulent fluctuations on SB evolution needs to be investigated in detail \citep{Chen_2020_ApJS}.

\subsection{Outline of the paper}
\label{sec:Intro_Outline}

Having reviewed previous observational results related to SBs and associated physical interpretations, we will now move to our analysis of SBs observed by \emph{PSP} during encounters 1 and 2. The intention of this work is not to verify or disprove either school of thought regarding SB origins described in Section \ref{sec:Intro_Origins}, but rather to analyze a set of observations that have the combined benefit of both being visually selected and also containing a large number of events, which will provide additional insight into both MHD and kinetic scale properties of SBs. This way, the results presented here provide a comprehensive set of  observational constraints for current, as well as developing future models of SB generation and evolution.

We use a database of visually selected SBs, with associated quiet and transition periods before and after each structure. For each event, we combine several methods to analyze plasma parameters, turbulent spectra, and intermittency levels. Additionally, as a proxy for turbulence behavior at ion scales, we compare levels of Stochastic Heating (SH), a non-linear plasma heating mechanism based on the breaking of magnetic moment by low-frequency turbulent fluctuations at ion scales \citep{Chandran_2010}, inside and outside of SBs. Details of these procedures are described in Section \ref{sec:Method}. In Section \ref{sec:Results}, we review our results and discuss possible implications. Finally, Section \ref{sec:Conclusion} summarizes the main points. 

% -------------------- METHOD -------------------- 

\section{Method} 
\label{sec:Method}

%%% SB database -------------------

\begin{figure}
\centering
\includegraphics[width=\textwidth]{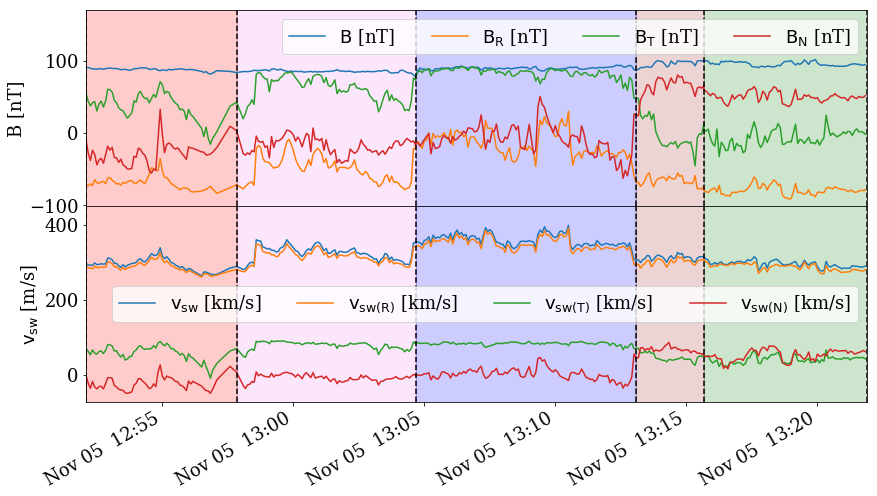}
\caption{Example of an event from Encounter 1, on Nov 5 2018, with five characteristic regions of interest --- from left to right: leading quiet region (LQR), leading transition region (LTR), switchback (SB), trailing transition region (TTR), and trailing quiet region (TQR), given in different colors.}
\label{fig:SB_Example}
\end{figure}

Within the heliospheric community, the exact definition and typical properties of SBs are still debated. Several different criteria to identify SBs in data sets have been introduced throughout the last few decades \citep{Yamauchi_2004_JGRA,Horbury_2018_MNRAS} and in particular in studies related to \emph{PSP} data \citep{DudokdeWit_2020_ApJS,Horbury_2020_ApJS,McManus_2020_ApJS,Mozer_2020_ApJS,Bourouaine_2020_ApjL}. Here, we use the database of SBs from \emph{PSP} Encounters 1 and 2 (E1 and E2) where, for each event, five separate regions are identified: 1) Leading Quiet Region (LQR) with stable velocities and magnetic fields before the SB; 2) Leading Transition Region (LTR), where the magnetic field rotates from LQR toward its SB orientation; 3) the SB itself with stable field orientation; 4) Trailing Transition Region (TTR); and 5) Trailing Quiet Region (TQR), with conditions which are, in general, not very different from the ones in LQR. The identification of events was done through visual inspections of $\mathbf{B}$ and $\mathbf{v_{sw}}$ time series. Magnetic field rotations coincident with the bulk velocity enhancement are selected, with the results being very consistent throughout multiple independent manual inspections. A total of 1,074 events\footnote{Further on, the term 'event' will be used for a sequence of the five different observed regions, while the term 'SB' will be used only for the period where the rotated magnetic field is observed.} were identified (662 in E1 and 412 in E2), 921 of which have all five distinct regions resolved. A typical example of a selected event is given on Figure \ref{fig:SB_Example}. For each of these regions separately, we extracted the average plasma parameters, including the proton core population density $n_p$, thermal velocity $v_t$ and solar wind bulk velocity $\mathbf{v_{sw}}$ from the Solar Probe Cup (SPC) measurements \citep{Kasper_2015,Case_2020_ApJS_SPC}, magnetic field $\mathbf{B}$ from the Fields instrument suite magnetometers \citep{Bale_2016} and temperatures perpendicular $T_\perp$ and parallel $T_{||}$ to $\mathbf{B}$ from analysis of effective temperatures measured by SPC and the fluctuating magnetic field \citep{Huang_2020_ApJS}. Only measurements flagged with the highest degree of confidence by the instrument teams are used in this study.

%%% turbulence fits -------------------

\begin{figure}
\centering
\includegraphics[width=\textwidth]{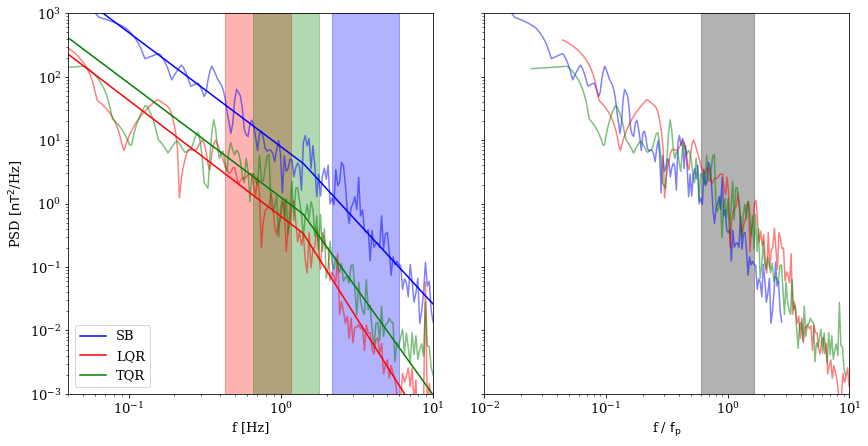}
\caption{Example of a power spectrum for LQR (red), SB (blue), and TQR (green) for a single event, with linear fits over the inertial and dissipation ranges. All three regions were sampled for $\sim$ 70 s. Shaded areas on the left panel represent the [$e^{-0.5} f_p$, $e^{0.5} f_p$] interval. The right panel shows the same three spectra with frequency normalized to $f_p$. }
\label{fig:SB_Fit}
\end{figure}

We also calculate the Fourier magnetic field trace Power Spectral Density (PSD) spectrum\footnote{For details on calculating PSD, see e.g. Appendix A of \cite{Bourouaine_2013}.} for each region, focusing on two characteristic regimes: 1) the inertial range where the Alfv\'enic fluctuations cascade toward smaller scales, characterized by a logarithmic scale slope usually measured in the domain [-1.7, -1.45] \citep{Bruno_2013_LRSP} and, after the spectral break frequency $f_b$,
2) the kinetic range where the fluctuations both become dispersive and the energy in the electromagnetic field is expected to more efficiently dissipate, eventually leading to plasma heating, causing significant steepening of the spectral slope \citep{Alexandrova_2012_ApJ,Verscharen_2019_LRSP}\footnote{This is a very simplified description of a typical turbulent spectrum. The border between the two described spectral regions is often observed as a third range at which the spectral steepening is associated with ion heating through kinetic processes \citep{Sahraoui_2010_PhRvL,Bowen_2020_PhRvL}.}. As multiple other phenomena, such as coherent structures likely generated by turbulence \citep{Lion_2016_ApJ,Howes_2016_ApJ}, magnetic reconnection \citep{Mallet_2017_JPlPh,Loureiro_2017_ApJ,Vech_2018b}, or instrument noise \citep{Martinovic_2019_ApJ} can interfere with measuring the background turbulent spectra, and as the event durations vary significantly, we calculated the spectral indices and the PSD at the ion scale break using a two step process. First, a Levenberg-Marquardt fit using a two line function, with the fitting range frequencies set manually assuring that both injection range and instruments noise are excluded, is performed for each LQR, SB and TQR region for a given event. This method provided results for the spectral break locations and PSD intensities with high accuracy and independent of initial guess and input parameters, but introduced a subjective bias on the values of inertial range logarithmic slopes through the manual selection of the low frequency limit. Therefore, the second step of the fitting algorithm was determining the inertial slopes by a least-square line fit of data in an automatically selected range. To avoid statistical uncertainties and the possible measurements of the injection range turbulence, the low frequency limit was set at 150\% of the minimum frequency of the power spectrum of a given region, while the high frequency limit was set at $0.75 f_b$, to avoid sampling the fine structure of the spectral break. An example of such a fit is given on Figure \ref{fig:SB_Fit}. We see both LQR and TQR spectra encounter the noise floor at $\sim$ 6 Hz. The frequency range contaminated by the noise is manually removed from the fitting procedure as to not affect the resulting slopes. In total, spectra from 1,863 regions (LQRs, SBs and TQRs) were fit successfully, and are included in this study.

It is worth emphasizing that a significant increase of $v_{sw}$ in SBs causes a shift of the spectrum in frequency space, in accordance with the shift prescribed by Taylor hypothesis \citep{Taylor_1938_RSPSA}. The \emph{Right panel} of Fig.~\ref{fig:SB_Fit} illustrates this property, with the three spectra, fairly different in terms of measured power, overlap as the observed frequency is normalized to $f_p$. Here, $f_p = v_{sw} \sin{\theta}/2\pi\rho_p$ is the convected gyrofrequency \citep{Bourouaine_2013}, $\theta$ is the angle between $\mathbf{v_{sw}}$ and \textbf{B}, $\rho_p = m_p v_{t\perp} / e_c B$ is the gyroradius, and $e_c$ is the elementary charge with all values averaged over a measured region. 

%%% PVI -------------------

As the magnetic field PSD does not capture the intermittent properties of turbulence, including phenomena such as localized coherent structures at the borders of the regions within each event, we estimate these properties using PVI, which is defined at time $t$ for some time increment $\tau$ as
\begin{equation}
    \mathrm{PVI} = \frac{|\Delta B (t, \tau)|}{\sqrt{\langle |\Delta B (t, \tau)|^2 \rangle}} 
\end{equation}
where $\Delta B = B(t) - B(t+\tau)$, $\langle\rangle$ denotes ensemble average over a region of interest (QR, TR or SB).
Based on previous studies, times with values of PVI $>$ 3 have been associated with non-Gaussian structures, events with PVI $>$ 6 categorized as current sheets, and with PVI $>$ 8 as reconnection sites \citep{Matthaeus_1982_JGR,Osman_2012,Greco_2018_SSRv,Chhiber_2020_ApJS}.

%%% SH -------------------

Detailed analysis of the turbulent spectra around $f_p$ provides estimates of ion heating delivered through non-linear mechanisms such as stochastic heating (SH). An overview of estimates for SH rates for the first two \emph{PSP} encounters is provided elsewhere \citep{Martinovic_2020_ApJS}, while in this study we will only focus on the comparison of SH in SBs and their associated quiet regions. The key quantity for estimation of SH is the normalized level of turbulent magnetic fluctuations at the convective gyroscale $\delta = \delta B / B$, leading to a total heating rate estimated as \citep{Hoppock_2019}
\begin{equation}
Q_\perp = v_A^2 \Omega \Bigg( \sigma_1 \delta^3 \exp \left[-\frac{\sigma_2}{\delta}\right] + \frac{\sigma^3 c_1 \delta^3}{\beta^{1/2}} \exp \left[-\frac{c_2 \beta^{1/2}}{ \sigma \delta}\right] \Bigg)
\label{eq:Heating_Rate}
\end{equation}
where $c_1 = 0.75$, $c_2 = 0.34$, $\sigma_1 = 5$ and $\sigma_2 = 0.21$ are order unity constants found from test particle simulations that are expected to be fairly insensitive to plasma parameters \citep{Chandran_2010,Hoppock_2019}. The relation between velocity and magnetic field fluctuations is quantified by the parameter $\sigma = 1.19$, proton cyclotron frequency is given as $\Omega = e_c B / m_p$, and $\beta = v_t^2 / v_A^2$. A detailed discussion of these constants is beyond the scope of this work, but can be found in previous literature on SH in the solar wind \citep{Chandran_2010,Xia_2013,Hoppock_2019,Martinovic_2019_ApJ}. The magnetic field fluctuation magnitude $P_B$ at the convective gyroscale is calculated by integrating the PSD in the vicinity of $f_p$, as shown on Figure \ref{fig:SB_Fit})
\begin{equation}
    \delta B = \Bigg[\frac{\pi}{C_0(n_{s})}  \int_{e^{-0.5} f_p}^{e^{0.5} f_p} P_B(f) df \Bigg]^{1/2} \label{eq:delta_B}.
\end{equation}
In Equation \ref{eq:delta_B}, $n_s$ is the PSD logarithmic slope in the integration range, determining the geometrical factor $C_0$, which is described in detail in previous studies \citep{Bourouaine_2013,Vech_2017}. We note that SH rates determined this way have very large uncertainties \citep{Martinovic_2019_ApJ}, but can be confidently used for comparison between different regions in a single event. 

% -------------------- RESULTS -------------------- 

\section{Results and Discussion} 
\label{sec:Results}

%%% macro parameters -------------------

\subsection{Overview of Plasma Parameters}
\label{sec:Results_MHD}

\begin{figure}
\centering
\includegraphics[width=\textwidth]{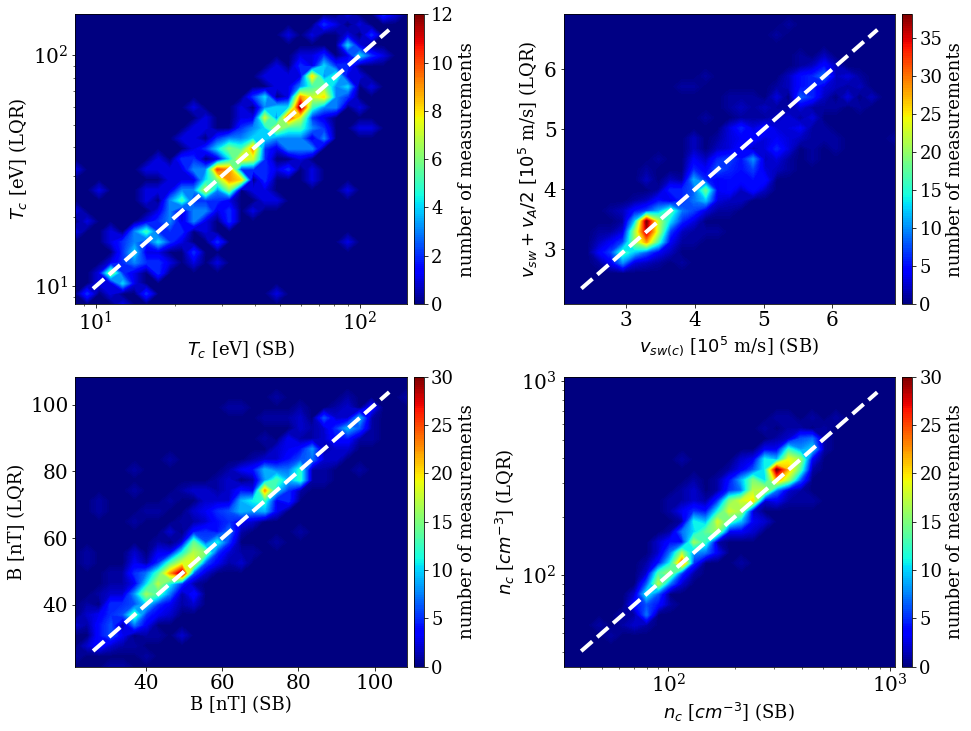}
\caption{Comparison of parameters in LQRs and SBs. The results for all the parameters are quite similar, with $v_{sw}$ being systematically increased in SBs by an average value of $\sim v_A/2$.}
\label{fig:SB_Comparison}
\end{figure}

The macroscopic parameters of SBs and LQRs are shown in Figure \ref{fig:SB_Comparison}. Magnetic field strength, core proton density and temperature\footnote{The SPC data set does not provide full values of temperature, but rather ones extracted from the reduced distributions. The scalar $T_c$ used here is provided by method described in \citep{Huang_2020_ApJS}.}, as well as other parameters which are not shown, are fairly similar inside and outside of SBs for the vast majority of events. The two peaks along the diagonal appearing in all three plots are due to the combination of data sets from E1 and E2, which have slightly different typical parameters. This is also the case for comparison between SBs and TQRs, not shown. These observations agree with the interpretation that plasma inside SBs is from the same folded plasma stream as the quiet regions outside the SB (see, e.g. \citep{Yamauchi_2004_JGRA,Tenerani_2020_ApJS,Fisk_2020ApJL,Wooley_2020_MNRAS}). However, it is not intuitively clear why the bulk velocity should increase by $\sim v_A/2$ within SBs. Even though the deflection angle\footnote{In this work, deflection angle is defined as the difference in average $\theta$ between LQR and SB for each event. The differences between TQR and SB are consistent to within several percent for large majority of events.} of the bulk velocity for SBs is very small \citep{Kasper_2019_Nature}, we find that the intensity of the $\Delta\mathbf{ v_{sw}}$ has a Gaussian distribution centered at $\sim 0.75 v_A$ (not shown). This value does not necessarily conflict with previous results, which argue that magnetic field deflections in SBs behave like almost purely Alfv\'enic phenomenon \citep{Yamauchi_2004_JGRA,Gosling_2011_ApJ,Matteini_2015_ApJ,Horbury_2018_MNRAS}, as these values are not comparing the states just before and after the transition regions, but are rather averaged over longer periods. One possible interpretation for such behavior is that the difference in bulk speeds, normalized to Alfv\'en speed, is proportional to the Alfv\'enic ratio between energy of magnetic field and velocity fluctuations, expected to be lower than unity in the solar wind \citep{Matteini_2014_GRL}, but this phenomena requires additional detailed investigation. 

%%% on turbulence properties -------------------

\subsection{Turbulence}
\label{sec:Results_Turbulence}

\begin{figure}
\centering
\includegraphics[width=\textwidth]{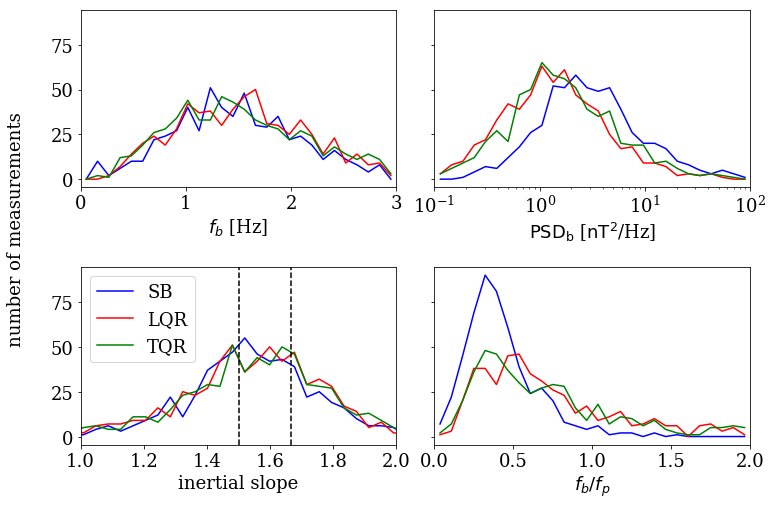}
\caption{Histograms for turbulent spectral properties measured in quiet regions and SBs, obtained from manually performed fits. \emph{Top panels} show that the ion scale break frequency in Hz $f_b$ remains unchanged in the shifted spectra of SBs, while the PSD at $f_b$ is enhanced due to the spectrum being shifted by the increase in $v_{sw}$. \emph{Bottom left panel} illustrates the fitted inertial range logarithmic slopes being similarly distributed in all three types of regions, and equally represented in the range between theoretically predicted values of -5/3 and -3/2. Finally, the \emph{bottom right panel} shows a notable shift in $f_p$ for SBs, affecting its ratio to $f_b$.}
\label{fig:SB_turbulence}
\end{figure}

Figure \ref{fig:SB_turbulence} shows the results from 1,863 manually fitted turbulent spectra in SBs and quiet regions. The most notable feature is that, even though the turbulent spectra are shifted in frequency due to increased $v_{sw}$, the spectral break from inertial to ion dissipation region spans the same frequency range. This implies the lowest wavenumber at which the dissipation is observed is lower in SBs, producing larger power at the break point (\emph{top panels}). \emph{Bottom right panel} shows comparison of the spectral break point with the convective gyroscale, also shifted toward higher frequencies (Figure \ref{fig:SB_Fit}), which is due to the increase in both $v_{sw}$ and $\theta$, given that all the other quantities comprising $f_p$ do not significantly change inside SBs (Figure \ref{fig:SB_Comparison}). The relation of the convected proton inertial length $f_d$ to $f_b$ (not shown) is very similar, which is expected as our data set does not contain any extremely high or low plasma $\beta$ measurements, as studied for example in \cite{Chen_2014_GeoRL}. It is important to note the contrast with recent results \citep{Duan_2020_ApJS} showing different scalings of $f_p$ and $f_b$ in PSP data; this difference is most likely masked in our results due to similarity in MHD plasma parameters in QRs and SBs (Figure \ref{fig:SB_Comparison}).

The fitted values of inertial range turbulence slopes shown in the \emph{bottom left panel} are consistent with two theoretically predicted regimes---a standard 'critically balanced' cascade with inertial slope centered around -5/3 \citep{Goldreich_1995}, and a 'dynamically aligned' cascade with the inertial slope of $\sim -3/2$ \citep{Boldyrev_2006_PhRvL}, describing a majority of the measured spectra with similar abundances in histograms. There is no preference to either of these regimes when comparing SBs to quiet regions. In order to understand the inertial slope behavior, we investigated possible relations of the fitted slope with multiple parameters, including plasma $\beta$, Velocity Distribution Function (VDF) moments, magnetic field and characteristics of its turbulent fluctuations, and radial distance. None of these parameters showed a correlation with the inertial slope values. However, this result does not necessarily contradict with previous studies where the inertial slope was observed to depend on the residual energy and intermittency \citep{Bowen_2018_ApJ}, cross helicity \citep{Chen_2013_ApJ}, radial distance \citep{Chen_2020_ApJS}, or proximity to the Heliospheric Current Sheet \citep{Chen_2021_AA}, due to relatively small ranges for these parameters measured in our data set. Similar studies to this work on the inertial slope dependencies will be preformed in the future as the \emph{PSP} data set expands, producing larger numbers of visually identified SB for analysis.

%%% about SH -------------------

\subsection{Comparison of SH Rates}
\label{sec:Results_SH}

To investigate properties of the ion scale turbulence and its interaction with VDF moments, we estimate the level of non-linear SH. Here, it is important to note that the SH values are very difficult to estimate for majority of events due to their relatively short duration and occasional strong wave activity. Therefore, these estimates should be observed only as a proxy for behavior of turbulence at the proton gyroscale, and its comparison inside and outside of SBs. The SH rates measured on this mission are considered in more detail and with more strict constraints elsewhere \citep{Martinovic_2020_ApJS}, while discussion of other solar wind heating mechanisms, such as Landau damping \citep{Quataert_1998,Chen_2019} or cyclotron heating \citep{Hollweg_1999JGR_1,Kasper_2013}, is beyond the scope of this paper.

\begin{figure}
\centering
\includegraphics[width=\textwidth]{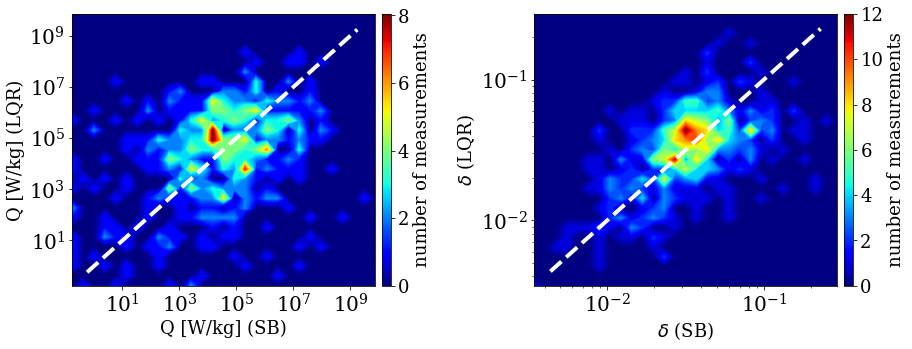}
\caption{Comparison of SH values (\emph{left}) and normalized magnetic field fluctuations at the convective gyroscale (\emph{right}) in LQRs and SBs. The values are very similar, as well as for the comparison between TQRs and SBs (not shown). The spread in the results originates from high uncertainties of quantities in Equations \ref{eq:Heating_Rate} and \ref{eq:delta_B}.}
\label{fig:SB_SH}
\end{figure}

The turbulent PSD at the convective gyroscale is related to the level of SH through the value of $\delta$, which is largely controlled by the amplitude of the PSD in the integration range around $f_p$, defined in Equation \ref{eq:delta_B}. This range is, for most cases, closer to $f_b$ in quiet regions than in SBs (Figure \ref{fig:SB_turbulence}). However, as the PSD at the break is significantly higher for SBs, the integration range levels, and consequently $\delta B / B$ at the gyroscale, remain very similar, as shown on Figure \ref{fig:SB_SH}. Considering that the parameters shown on Figure \ref{fig:SB_Comparison} are also very similar, the value of SH in SBs remains approximately the same and thus the existence of SBs is not expected to enhance the contribution of SH to solar wind heating. Observing Figure \ref{fig:SB_Fit}, we could argue that the similarity in the estimated SH values originates from the wave vector PSD, Doppler shifted to different frequencies. However, two important characteristics must be taken into account when considering SH rates --- 1) the PSD at $f_b$ is increased for SBs and 2) the frequency of interest $f_p$ does not represent a standard Doppler shift, but is also affected by the angle $\theta$, pushing the integration range to slightly higher frequencies. These two factors have effects that partially negate one another, providing to the level of accuracy available, similar values of SH rates.

\begin{figure}
\centering
\includegraphics[width=0.8\textwidth]{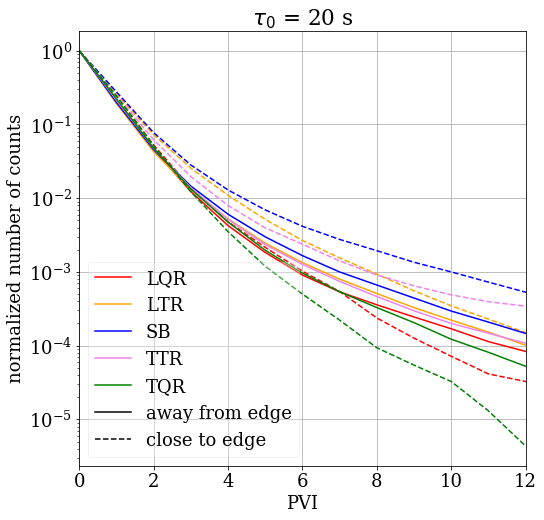}
\caption{PVI histograms averaged for different regions. Solid lines represent the measurements made more than $\tau_0 = 20$ s from edge of each respective region, while dashed lines are from the data sampled within $\tau_0$ time from each edge.}
\label{fig:SB_PVI}
\end{figure}

%%% PVI and intermittency ------------------------

\subsection{Effects of Borders and Intermittency}
\label{sec:Results_Intermittency}

Similar characteristics of the turbulent spectra shown in Figure \ref{fig:SB_turbulence} do not reflect the potentially different intermittency levels in different regions. The distribution of PVI values, used as a proxy for intermittency, is shown on Figure \ref{fig:SB_PVI} organized by the five kinds of regions, clearly demonstrating that PVI is the highest for the case of SBs, then TRs, and is the lowest for QRs \citep{Huang_2020_A&A}. As recent results \citep{Krasnoselskikh_2020_ApJ,Agapitov_2020_ApJL,Farrell_2020_ApJS} have shown that the edges of SBs can be populated by various structures and exhibit strong wave activity, we separately observe the measurements close to the edges of each region. Solid lines are calculated using only PVI values measured at least $\tau_0 = 20$ s from the region's edge. The histograms start to deviate for SB and transition regions for PVI = 3-6, with the difference of $\sim 10\%$ at PVI $\approx 3$ and for about a factor of 2.4 at PVI $\approx 10$. On the other hand, the magnetic field sampled close to the region edges are featuring significantly more drastic trends, with spread of a factor of 2.3 for PVI $\approx 3$, and over an order of magnitude for PVI $\approx 10$. This result is insensitive to the $\tau_0$ used; recalculating Figure \ref{fig:SB_PVI} with $\tau_0$ of 1, 2, 5, 10, and 50 s (not shown) yields very similar results. We also reproduced Figure \ref{fig:SB_PVI}, but only taking into account periods either larger or smaller than various time durations between 2 and 200 s. The results are very similar to the ones shown on the Figure, and are not displayed here. From here, we conclude that properties of PVI are independent of the region duration. This result is fully consistent with previous observations findings of SBs having increased intermittency \citep{Horbury_2018_MNRAS}, as well as the presence of waves and enhanced currents at the edges of SBs. The presence of these structures at the SB borders suggests that they might be responsible for increased 'lifetime' of SBs, stabilizing these structures against decay \citep{Landi_2005_ApJ}.

%%% SH and intermittency ------------------------

\begin{figure}
\centering
\includegraphics[width=\textwidth]{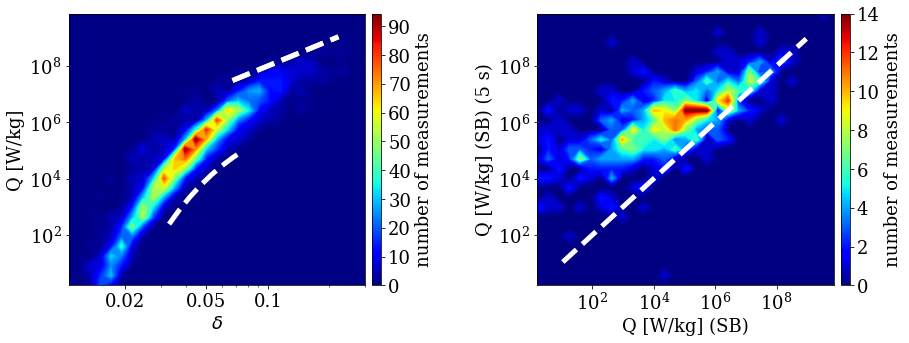}
\caption{Effects of intermittency on SH rates. \emph{Left panel} illustrates the major importance of the exponential factors in Equation \ref{eq:Heating_Rate} for weaker SH, and dominance of $\delta^3$ term for strong SH. \emph{Right panel} shows that the effects of intermittency---illustrated through average SH rate of 5 s intervals inside a SB compared to SH rate calculated for the entire region---are notably more important when the exponential factor is able to significantly modify the fluctuation damping rate \citep{Mallet_2019_JPP}.}
\label{fig:SB_SH_int}
\end{figure}

Finally, it is interesting to consider the effects of intermittency on SH. Figure \ref{fig:SB_SH_int} illustrates the SH values calculated for every 5 s interval and then averaged over the entire region. It is notable that the intermittency strongly enhances the level of SH whenever that level is low, but it is not the case for intervals with strong SH, which are responsible for the majority of SH's contribution to solar wind heating \citep{Martinovic_2019_ApJ,Martinovic_2020_ApJS}. This feature, shown on \emph{right panel} of Figure \ref{fig:SB_SH_int}, was verified to also hold for quiet regions and any time averages in [5-100] s range. To understand this effect, we refer to the work of \cite{Mallet_2019_JPP}, who calculated fluctuations damping rates $\gamma \sim Q / \delta z^2$ for different solar wind heating mechanisms. Here, $Q$ is the theoretical heating rate, and $z$ is the Elsasser variable \citep{Elsasser_1950_PhRv} at a given scale. For the case of Landau damping, intermittency has no effect on the heating intensity due to the damping rate increasing linearly with $\delta z$. On the other hand, the SH was expected to be strongly affected by intermittency due to the exponential factor in Equation \ref{eq:Heating_Rate}. However, the \emph{left panel} shows that SH dependence on $\delta$ transitions from an exponential to a power law function. This implies that the damping rate for the case of SH is approaching the linear function as the exponential factor either approaches unity, becoming a very weak function of $\delta$. Therefore, when SH is a very important, and possibly dominant, heating mechanism, its level and variance within SBs and quiet regions, are significantly less affected by intermittency compared to intervals when SH is low. 

However, another level of analysis of this topic is related to the relation between intermittency and order unity constants $c_1$, $c_2$, $\sigma_1$ and $\sigma_2$, given in Equation \ref{eq:Heating_Rate}. Results of RMHD simulations \citep{Xia_2013} indicate that their predicted values can be notably decreased depending on the nature of the turbulent fluctuations, increasing the predicted SH rate. This topic requires separate detailed investigation which is outside of the scope of this paper.

%%% Conclusions ------------------------

\section{Conclusion}
\label{sec:Conclusion}

In this work, a comprehensive analysis of manually selected magnetic field SB events in \emph{PSP} encounters 1 and 2 is performed in order to determine the characteristics of plasma inside and outside of the regions where the field is deflected. The investigation shows that the plasma has very similar bulk quantities and magnetic field values inside SBs, compared to the equivalent values in the quiet regions before and after a SB, with the very well known shift in the solar wind bulk speed being of the order of the Alfv\'en speed. Additionally, estimates of the non-linear SH rates inside and outside SBs show that the heating rate due to this mechanism remains fairly similar, which suggests that the plasma sampled by the spacecraft in different regions of a single event most likely belong to a single plasma stream.

However, several questions are raised related to the turbulent cascade and kinetic processes in these events. First, the magnetic field power spectra ion-scale breaking frequency $f_b$ remains similar inside and outside of SBs, in spite of the spatial turbulent spectra in SBs being shifted toward higher frequencies due to effects of Taylor hypothesis. Second, we observed the inertial range PSD logarithmic slopes to have a distribution consistent with both the values of -5/3 and -3/2, characteristic for the classic critical balance and dynamic alignment turbulence theories, respectively. The inertial logarithmic slope is not sensitive to typical physical quantities such as density, pressures, or radial distance, and its phenomenology remains unclear, but this could be due to relatively limited range of the plasma parameters measured. Third, the increased values of PVI in the transition regions and SBs compared to quiet regions, with this property being independent of the region duration, reflects the existence of small scale structures at the region borders. The origin of these structures is directly related to the SB origins and will be one of the central questions regarding SB evolution in need of resolution.

In reference to the two different schools of thought about SB origins discussed in Section \ref{sec:Intro_Origins}, the fact that we generally observe the same plasma inside and outside of what is most likely an S-shaped structure \citep{DudokdeWit_2020_ApJS,Horbury_2020_ApJS}, implies that either 1) a parcel of solar wind plasma expands outwards until the turbulent fluctuations become sufficiently strong to fold magnetic field back on itself, forming a SB; or 2) the magnetic field structure travels through the solar wind with the relative speed of the order of $v_A$, folding the plasma without significantly altering its internal structure. In the second scenario, complicated processes near the solar surface, such as interchange reconnection (see e.g. \cite{Fisk_2020ApJL,Zank_2020_ApJ}), create a structure that persists as it propagates through the solar wind, temporarily increasing the speed of the part of the stream which is being pulled through the structure.

The transition regions and edges themselves are not investigated in detail in this work. Although the basic phenomenology of these regions is well understood, their comparison with quiet regions and SBs is not trivial, and the magnetic field PSD from these regions require a more sophisticated interpretation. The kinetic scale processes and ion-scale turbulence inside SBs, and development and dissipation of current sheets at their edges, remain an open topic for future effort. Finally, we note that all the remarks made in this paper are meant to describe characteristics of the majority of observed events, and therefore provide observational constraints for models aiming to describe SBs. We do not expect future case studies, which would cover single or few interesting events in more detail, to necessarily replicate all the general conclusions given here. 

\acknowledgments

The SWEAP Investigation and this publication are supported by the \emph{PSP} mission under NASA contract NNN06AA01C.
The FIELDS experiment was developed and is operated under NASA contract NNN06AA01C. 
M. M. Martinovi\'c and K. G. Klein were financially supported by NASA grants 80NSSC19K1390 and 80NSSC19K0829. B. D. G. Chandran is partially supported from NASA grants NNX17AI18G and 80NSSC19K0829 and NASA grant NNN06AA01C to the Parker Solar Probe FIELDS Experiment.
E. Lichko is supported by the National Science Foundation under Award No. 1949802. C. H. K. Chen was supported by STFC Ernest Rutherford Fellowship ST/N003748/2 and STFC Consolidated Grant ST/T00018X/1.

%\bibliographystyle{aasjournal}
%\bibliography{Latex_Refs}

\end{document}